\documentclass[]{emulateapj}

\def\rmit#1{{\it #1}}              

\def\ie{\rmit{i.e.}}
\def\eg{\rmit{e.g.}}

\usepackage{natbib,graphicx}

\usepackage{color}

\shorttitle{Holography of a sunspot}

\shortauthors{Felipe et al.}

\begin{document}

\title{Helioseismic holography of simulated sunspots: magnetic and thermal contributions to travel times}

\author{T. Felipe\altaffilmark{1,2,3}, D. C. Braun\altaffilmark{3}, A. D. Crouch\altaffilmark{3}, A. C. Birch\altaffilmark{4} }
\email{tobias@iac.es}

\altaffiltext{1}{Departamento de Astrof\'{\i}sica, Universidad de La Laguna, 38205, La Laguna, Tenerife, Spain} 
\altaffiltext{2}{Instituto de Astrof\'{\i}sica de Canarias, 38205,
C/ V\'{\i}a L{\'a}ctea, s/n, La Laguna, Tenerife, Spain}
\altaffiltext{3}{NorthWest Research Associates, Colorado Research Associates, Boulder, CO 80301, USA}
\altaffiltext{4}{Max-Planck-Institut f\"{u}r Sonnensystemforschung, Justus-von-Liebig-Weg 3, 37077 Göttingen, Germany}

\begin{abstract}

Wave propagation through sunspots involves conversion between waves of acoustic and magnetic character.  In addition, the thermal structure of sunspots is very different than that of the quiet Sun.  As a consequence, the interpretation of local helioseismic measurements of sunspots has long been a challenge.  With the aim of understanding these measurements, we carry out numerical simulations of wave propagation through sunspots.   Helioseismic holography measurements made from the resulting simulated wavefields show qualitative agreement with observations of real sunspots.   We use additional numerical experiments to determine, separately, the influence of the thermal structure of the sunspot and the direct effect of the sunspot magnetic field.   We use the ray approximation to show that the travel-time shifts in the thermal (non-magnetic) sunspot model are primarily produced by changes in the wave path due to the Wilson depression rather than variations in the wave speed.  This shows that inversions for the subsurface structure of sunspots must account for local changes in the density.  In some ranges of horizontal phase speed and frequency there is agreement (within the noise level in the simulations) between the travel times measured in the full magnetic sunspot model and the thermal model. If this conclusion proves to be robust for a wide range of models, it would suggest a path towards inversions for sunspot structure.

\end{abstract}



\section{Introduction}

Local helioseismology probes the solar interior by analyzing the wave field observed at the surface. Several different techniques have been developed over the years, including Fourier-Hankel analysis \citep{Braun+etal1987}, ring-diagram analysis \citep{Hill1988}, time-distance helioseismology \citep{Duvall+etal1993}, and helioseismic holography \citep{Lindsey+Braun1990}. Sunspots and active regions have been prominent targets of these studies, but still a clear picture of the Sun's subsurface magnetic activity and its relation with helioseismic signals remains elusive. 

Early results using Fourier-Hankel analysis showed that sunspots can absorb up to half of the incident acoustic wave power and shift the phase of the waves \citep{Braun+etal1988, Braun1995}. This work was followed by several theoretical attempts to explain the observed absorption, with mode conversion as the most promising candidate \citep{Cally+Bogdan1993, Cally+etal1994, Crouch+Cally2003, Cally+etal2003}. This shows the relevance of including wave interactions with magnetized atmospheres in the modeling of active regions. 

Inversion procedures were subsequently developed to infer the subsurface wave-speed of sunspots \citep{Kosovichev1996, Kosovichev+etal2000} from $p-$mode travel-time shifts, defined as the difference in the travel time between the sunspot observations and those expected for the quiet Sun. In their methodology, the effect of the magnetic field is indirectly accounted for by the modification that it produces in the wave speed of the medium. Since the travel-time shift changes with the phase speed of the waves, from longer travel times at small phase speed (shallow modes) to shorter travel times at higher phase speed (deep modes), inversions produce a two-layer subsurface model with a negative wave-speed perturbation in the top 4--5 Mm layer and a positive perturbation at depths between 5 and 10 Mm below the surface \citep{Couvidat+etal2006}.

\begin{figure*}[!ht] 
 \centering
 \includegraphics[width=18cm]{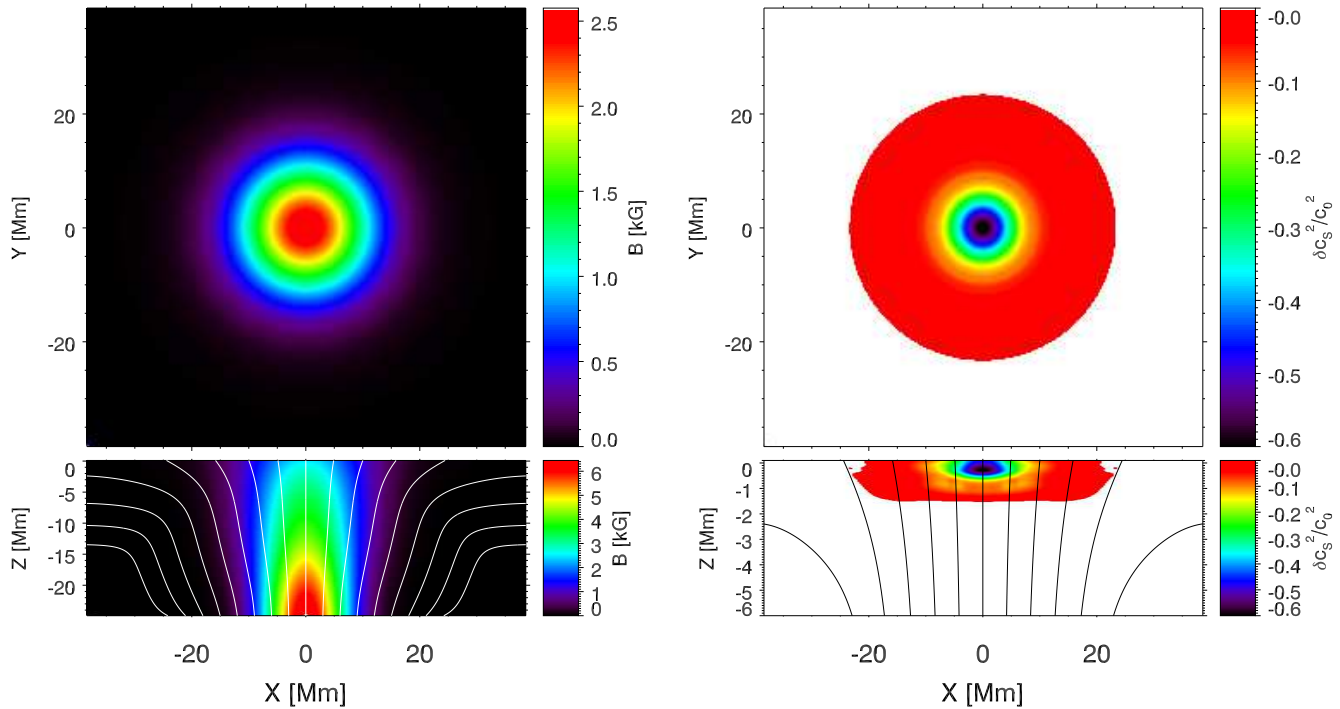}
  \caption{Properties of the sunspot model atmosphere. Left panels: magnetic field; right panels: square of the perturbation of the sound speed. Top panels: horizontal plane at the photosphere; bottom panels: vertical plane at y=0. The lines in the bottom panels represent magnetic field lines. Note that the bottom right panel represents a near-surface region.}
  \label{fig:sunspot_model}
\end{figure*}

In the last decade many studies have suggested that magnetic regions can produce disproportionately large perturbations to travel times within the first Mm or less below the photosphere. \citet{Lindsey+Braun2005} discussed the so-called ``showerglass'' effect, which produces phase distortions that contribute to the helioseismic signal, while \citet{Schunker+etal2005} showed evidence of the variations of the helioseismic signature with the line of sight in inclined magnetic fields. Using helioseismic holography, \citet{Braun+Birch2006,Braun+Birch2008} found a strong frequency dependence in observed travel-time shifts in sunspots, including changes of signs, and proposed that it might be produced by near-surface changes in wave propagation properties. A similar dependence was found by \citet{Couvidat+Rajaguru2007}, who also found ring-shaped sound-speed increases in their inversions. These were interpreted as artifacts arising from a failure in the inversion procedure to account for surface effects of the magnetic field of the sunspots.

Traditionally, helioseismic data have been interpreted in terms of variations of the isotropic sound speed of the medium and flows below the surface. However, new evidence points to the relevance of the direct effect of magnetic fields in those measurements \citep{Cameron+etal2008,Cally2009,Moradi+etal2009}. Mode conversion is one of the processes that can introduce phase shifts. Fast magnetic waves are generated from the conversion of acoustic waves at the region where the sound and Alfv\'en speed are similar \citep{Schunker+Cally2006}. These waves are reflected back to the photosphere due to the gradient in the Alfv\'en speed \citep{Khomenko+Collados2006}, or are partially converted into upgoing and downgoing Alfv\'en waves \citep{Hansen+Cally2012, Khomenko+Cally2012, Felipe2012}. Downward propagating fast and Alfv\'en waves may leave a trace on the travel-time perturbations measured from local helioseismology techniques \citep{Cally+Moradi2013}. Accurate modeling of wave interaction with magnetic fields and its influence in the helioseismic signal is strongly needed. This development should lead to the inclusion of the physical effects of the magnetic field when inverting for the subsurface structure of sunspots. 

In this context, numerical simulations are a promising way to understand wave propagation through magnetic fields and to test local helioseismic techniques. Several attempts have been made in the last years. \citet{Birch+etal2009} studied several hydrostatic models with regions of enhanced sound speed. The shallow case qualitatively reproduced the observed travel-time variations with frequency, supporting the suggestion of \citet{Braun+Birch2006}. The simulations of wave propagation in a magnetohydrostatic (MHS) model by \citet{Moradi+etal2009} also reproduced this behavior. Later, \citet{Birch+etal2011} performed some tests of helioseismic holography using two cases of non-magnetic models with the wave-speed structures inferred from observations, including the two-layer model \citep{Kosovichev+etal2000} and a one-layer model with a near-surface increase of the sound speed \citep{Fan+etal1995}. Their inversions were able to qualitatively recover the structure of the background models. In a recent work, \citet{Braun+etal2012} compared travel-time shifts measured from a realistic magnetoconvective sunspot simulation with those obtained from two actual sunspots. The results show remarkable agreement and stress the inability of a perturbation to the sound speed to account for the observed travel-time shifts.

The aim of this work is to analyze the physical causes of the travel-time perturbations measured in sunspots. To this end, we have developed numerical simulations of wave propagation in sunspot models. The organization of the paper is as follows: numerical simulations are described in Section \ref{sect:numerical}, travel-time signals measured using helioseismic holography are shown in Section \ref{sect:holography}, and these results are interpreted with the help of calculations in the ray approximation in Section \ref{sect:rays}. Finally, discussion and conclusions are presented in Section \ref{sect:conclusions}.

\section{Numerical simulations}
\label{sect:numerical}

We used the code MANCHA \citep{Khomenko+Collados2006, Felipe+etal2010a} to solve the three dimensional magnetohydrodynamic equations. The code calculates the evolution in a background model of a perturbation driven by a force added to the equations. A Perfect Matched Layer \citep{Berenger1996} is placed at the top and bottom boundaries in order to absorb waves without reflections, and the boundary conditions are periodic in the horizontal directions. In this paper we present several simulations, all of them with the same computational domain with dimensions 102.4 $\times$ 102.4 $\times$ 26 Mm$^3$ with a resolution of 0.2 Mm in the horizontal directions and 0.05 Mm in the vertical direction, using a grid with 512 $\times$ 512 $\times$ 520 cells. The bottom boundary is located at $z=-25$ Mm and the top boundary at $z=1$ Mm, with $z=0$ Mm corresponding to the photospheric level, where the optical depth at 500 nm is unity in the quiet Sun atmosphere. In order to reduce the strong constraint of the time step imposed by the high Alfv\'en velocity in low $\beta$ regions, we have limited the strength of the Lorentz force following \citet{Rempel+etal2009}. The maximum Alfv\'en velocity has been set to 80 km s$^{-1}$ \citep[see][for a discussion of the influence of a Lorentz force limiter in helioseismic travel-times]{Moradi+Cally2014}.

The background model was obtained following the \citet{Khomenko+Collados2008} method. It produces thick flux tubes with distributed currents. The flux tubes are azimuthally symmetric and show no twist. The thermodynamic variables far from the axis of the sunspots are adopted from CSM\_B model from \citet{Schunker+etal2011}, which consists of a quiet Sun model stable against convection obtained by modifying the vertical pressure gradient of Model S \citep{Christensen-Dalsgaard+etal1996}. The atmosphere between the quiet Sun boundary and the magnetized atmosphere at the axis of the spot merges smoothly. The sunspot model has a photospheric magnetic field strength of 2500 G and a Wilson depression of 450 km. The coefficient of specific heats $\Gamma_1$ is obtained from the OPAL equation of state \citep{Rogers+etal1996} using the abundances of the standard model. It is a function of height $z$ and radial position $r$. Figure \ref{fig:sunspot_model} shows the magnetic field (left panels) and square of the perturbation of the sound speed (right panels), obtained as $\delta c^2(r,z)= c^2(r,z)-c_0^2(z)$, where $c(r,z)$ corresponds to the sound speed of the sunspot model and $c_0(z)$ is the sound speed in the quiet Sun atmosphere. $\delta c^2(r,z)$ consists of a near-surface reduction of around 60\% of the quiet Sun sound speed.   

In this work we also performed other two numerical experiments. In the first, we computed the wave field using as a background model the thermal structure of the sunspot, but setting the magnetic field to zero throughout the domain. In the following we will refer to this case as the ``thermal spot''. In the other case, we used the thermal structure of the quiet Sun (without horizontal variations) but introduced the magnetic field of the sunspot. This case will be called the ``magnetic-only spot''. This way, we can isolate the thermal and magnetic components of the sunspot, and independently evaluate their effect on the travel-time measurements. A quiet-Sun simulation was also computed in order to apply the method of noise subtraction \citep{Werne+etal2004}. In our simulations the main source of noise is realization noise due to the stochastic nature of the driver (see next paragraph). As suggested by \citet{Werne+etal2004}, this noise can be estimated by performing a simulation without the perturbation in the background model but using the same source excitation. Consequently a considerable reduction in the noise can be achieved by taking the difference between the sunspot travel-time shifts and those measured from the simulation without background perturbation (a quiet Sun simulation in our case). This technique has been used in several prior studies \citep[\ie][]{Hanasoge+etal2007, Birch+etal2009,Dombroski+etal2013}.

The MANCHA code solves the MHD equations for perturbations. The equilibrium state is explicitly removed from the system of equations. In the full sunspot case, the background model follows the equation of MHS equilibrium and the solution is physically accurate. For the thermal and magnetic-only sunspots, the terms of the MHS equation were also removed, even though in these cases they do not cancel out. The resulting equations, which describe the wave propagation, do not require pressure balance in the background state. In addition, the equations have been linearized by neglecting second and higher-order terms and the wave amplitude has been restricted to the range where the linear approximation holds. The variations of all atmospheric parameters associated with the wave field have arbitrarily small amplitudes. As a result, the effect of the wave pressure disturbance on the pressure-unbalanced model is negligible. This strategy allows us to determine the independent contributions of the thermal and magnetic perturbations to the travel-time shifts. \citet{Moradi+etal2009, Moradi+etal2015} have successfully employed a similar approach for suppressing the direct magnetic effect on the waves (thermal sunspot) by means of linear numerical simulations.

\begin{figure}[!ht] 
 \centering
 \includegraphics[width=9cm]{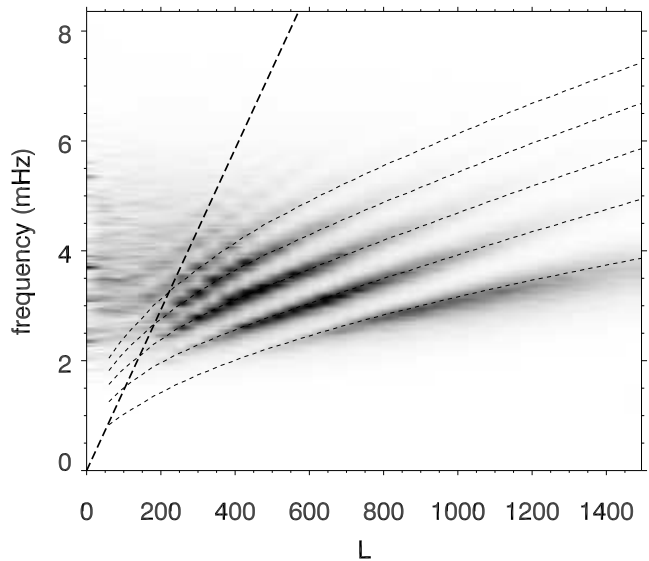}
  \caption{Power spectra of the photospheric vertical velocity from the quiet-Sun simulation. Black indicates regions of high power and white low power. The black short-dashed curves represent the eigenfrequencies of Model S and the long-dashed line indicates where the phase speed equals the sound speed at the bottom of the simulation.}
  \label{fig:power_spectra}
\end{figure}

The wave field is excited by sources added in the equations, each of them with the spatial and temporal behavior described by \citet{Parchevsky+etal2008}. They are randomly distributed in the horizontal directions. In the quiet-Sun region of the model the vertical position of the sources is $z=-0.15$ Mm, while their depth is increased for the sources closer to the center of the sunspot following a constant temperature surface in order to mimic the Wilson depression. Note that we have imposed the same set of sources for all the computations (based on the thermodynamic structure of the sunspot model), even for the quiet Sun simulation and the ``magnetic-only spot'', despite the lack of a horizontal variation of the temperature in these cases. A new source starts every time step. This driver produces a wave spectrum which resembles the solar spectrum (Figure~\ref{fig:power_spectra}). The duration of all the simulations is 8 hr, and the output is saved with a cadence of 45 seconds. The analyses presented in this paper are based on the vertical velocity at the geometrical height given by a surface with constant optical depth $\tau=0.01$ from the sunspot model. Note that the magnetic-only sunspot does not have a Wilson depression, since its thermal structure at all spatial positions is the same as in the quiet Sun model. Even in this case, we selected the vertical velocity at the same geometrical depth used in the other two models in order to make the measurements more comparable.

Figure \ref{fig:power_spectra} shows the power spectra obtained for the vertical velocity of the quiet Sun simulation at the photosphere. The model S eigenfrequencies \citep[obtained following][]{Birch+etal2004} are plotted as a reference. The shift of the power from the location of the ridges in Model S is produced by the modification of the vertical pressure gradient of CSM\_B model in order stabilize the atmosphere against convection \citep{Schunker+etal2011}. The straight long-dashed line indicates the region of the $k-\omega$ domain with a horizontal phase speed of 64 km s$^{-1}$, corresponding to the speed of sound at the bottom boundary (z=-25 Mm). Waves with higher phase speed reach the bottom of the domain and are damped by the PML layer, so most of the power at the left of the line is suppressed. 

Some spurious reflections may reflect back from the top boundary. However, the effect of these spurious reflections on the measurements is negligible. First, their amplitude is strongly damped at the PML and the contribution of their signal is very low. Second, fast magnetoacoustic waves in the low-beta atmosphere are reflected back towards the interior due to the gradients in the Alfven speed \citep{Rosenthal+etal2002, Khomenko+Collados2006}, so most of them do not reach the top boundary. Finally, acoustic waves are reflected at the height where their frequency equals the cut-off frequency. In our sunspot model, the maximum of the cut-off frequency at all spatial positions is above 5.7 mHz. In this work, we have analyzed the travel times for waves with frequencies below 5.25 mHz, so we are mainly restricted to waves trapped in the solar interior. Thus, they do not hit the top boundary. Only at those locations where the magnetic field is significantly inclined, slow magnetoacoustic waves can travel to the high atmospheric layers due to the reduction of the cut-off frequency with the cosine of the inclination angle \citep{Jefferies+etal2006}. In these regions with inclined magnetic field the fast-to-fast mode conversion is more efficient \citep{Cally2005}. Most of the wave power at the low-$\beta$ atmosphere (for the frequencies of interest) will be carried by the fast magnetoacoustic wave and, as discussed previously, they will be naturally reflected back towards the interior. All in all, spurious reflections from the top boundary have a negligible impact on the travel time measurements.

\section{Helioseismic holography}
\label{sect:holography}

\subsection{Procedures}
\label{sect:procedures}

The travel-time maps were measured following the general procedures for surface-focused helioseismic holography. This method is described in \citet{Braun+Birch2008} and has been applied for the analysis of synthetic data presented in previous works using sound speed perturbations \citep{Birch+etal2009, Birch+etal2011} or realistic magnetoconvective simulations \citep{Braun+etal2012}. Helioseismic holography estimates the wave field at ``focus points'', located at a chosen depth and position in the solar interior, assuming that the observed wavefield at the surface in a region called the ``pupil'' is produced by waves diverging from the focus point or wave converging toward that point. These quantities, called the egression and ingression, are obtained from the convolution of the surface oscillatory signal with appropriate Green's functions. Green's functions can be interpreted as propagators between the focus point and the surface wavefield. Local control correlations are calculated as the correlation between the surface observed wavefield and the ingression/egression at the surface \citep[Equations 1--2 from][]{Braun+Birch2008}. They provide the phase shift of the incoming and outgoing waves relative to the phase of the same waves propagating in the solar model used to compute the Green's functions. In the following we will be interested in the travel-time shifts, which are obtained from the phase of the local control correlations \citep[Equation 3 from][]{Braun+Birch2008}.

The first step in the data analysis consists of multiplying the Fourier transform of the photospheric vertical velocity extracted from the simulation with a chosen filter. We have used phase-speed filters which isolate waves with a range of horizontal phase speed described in Table 1 from \citet{Couvidat+etal2006}. We used the filters TD1 (central phase speed of 12.8 km s$^{-1}$) through TD5 (central phase speed of 35.5 km s$^{-1}$). A specific pupil function is employed with each filter, as given by the same table. 


Then, local control correlations are measured. They are analogous to center-annulus cross-covariances used in time-distance helioseismology. As a next step, we apply filters in the temporal frequency centered at 2.75, 3.25, 3.75, 4.25, 4.75, and 5.25 mHz with bandpass widths equal to 0.5 mHz. Travel-time shifts are measured from the filtered correlations. Finally, the noise-corrected results are obtained by subtracting quiet Sun travel-time shifts from the sunspot travel-time shifts.

\begin{figure}[!ht] 
 \centering
 \includegraphics[width=9cm]{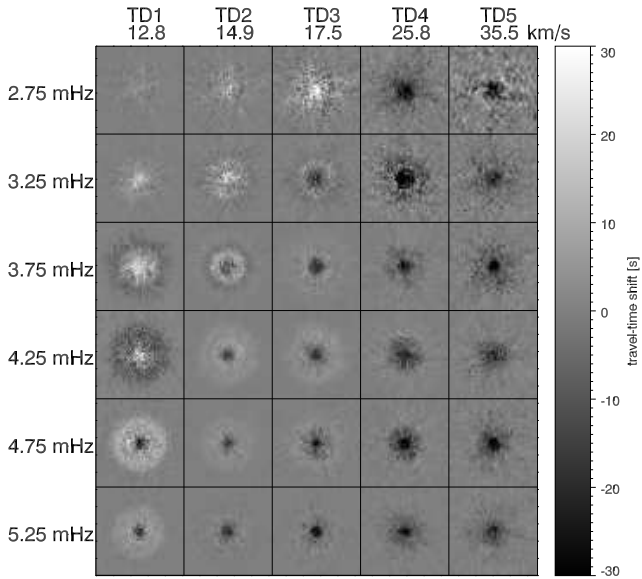}
  \caption{Mean travel-time shifts measured for the full sunspot simulation using phase speed and frequency bandpass filters. Columns correspond to the different the phase speed filters used, with the name of the filter and its phase speed (in km s$^{-1}$) indicated at the top. Each row shows a frequency filter, with the central frequency of the bandpass indicated at the left side.}
  \label{fig:sunspot travel-time phase}
\end{figure}

\subsection{Sunspot travel-time maps}
\label{sect:sunspot travel-time}

Figure \ref{fig:sunspot travel-time phase} shows phase-speed filtered mean travel-time maps of the full sunspot simulation. They are computed as the average of the incoming and outgoing travel-time shifts relative to the quiet Sun reference simulation. The travel times show a qualitative agreement with previous observational and numerical data \citep{Braun+Birch2008, Braun+etal2012}: low phase speed waves with lower frequencies present positive travel-time shifts, while as higher frequency and higher phase speed filters are considered the travel-times are progressively reduced, including a change in their sign. The phase-speed filter TD1 shows positive travel-time shifts (around 20 s) for frequency filters centered between 3.25 and 4.25 mHz. Negative travel-time shifts are observed at the center of the sunspot for higher frequencies, but they are surrounded by a ring of positive travel-time shifts. The change in sign of the travel-time shift appears at 3.75 mHz for the TD2 filter and 3.25 mHz for the TD3 filter. Higher phase-speed filters show negative travel-time shifts for all frequencies. 

Travel-time shifts between -20 and -30 s are found for most of the combinations of phase speed and frequency filters that present a negative travel-time. These perturbations are smaller than those found in realistic magnetoconvective simulations \citep{Braun+etal2012}, but it should be noted that the properties of the MHS sunspot model used for our simulations (magnetic field strength, radial size, and Wilson depression) are different from those of the simulation analyzed in \citet{Braun+etal2012}. Quantitative differences can also be identified with respect to observational results \citep{Braun+Birch2008, Braun+etal2012}. Since our model was not designed to reproduce any specific set of numerical or observational data, we consider the qualitative agreement to be highly significant, and we are confident that our simulation of wave propagation on a MHS sunspot model captures most of the relevant physics to study this problem.

\begin{figure}[!ht] 
 \centering
 \includegraphics[width=9cm]{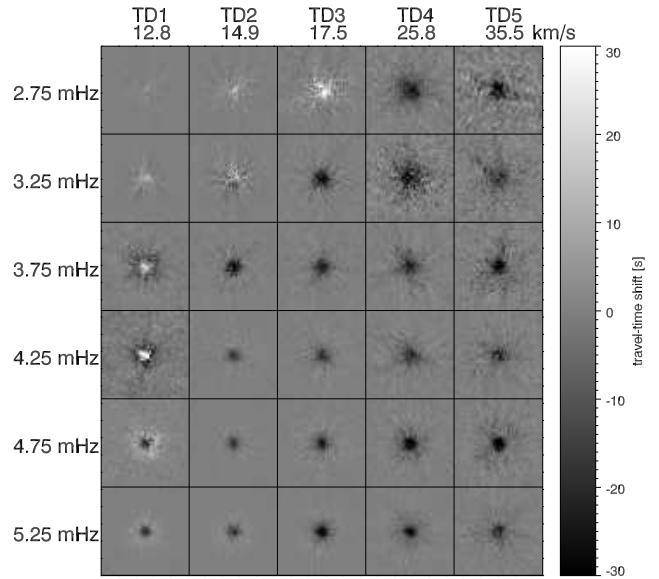}
  \caption{Mean travel-time shifts measured for the thermal sunspot simulation using phase speed and frequency bandpass filters, as in Figure \ref{fig:sunspot travel-time phase}.}
  \label{fig:thermal travel-time phase}
\end{figure}

\subsection{Thermal spot travel-time maps}
\label{sect:thermal travel-time}

In this simulation, the atmospheric model has the same pressure, density and $\Gamma_1$ from the sunspot model, but the magnetic field is set to zero. Figure \ref{fig:thermal travel-time phase} shows the travel-time maps obtained for this numerical simulation. These measurements isolate the influence of the indirect effects of the magnetic field (\ie, the modification that the magnetic field produces in the stratification of the thermodynamic variables) on the travel-time shifts. The travel-time map for TD1 and frequency 4.25 mHz has an abnormal behavior, since it exhibits a negative travel-time perturbation of about seven seconds at quiet Sun regions far from the center of the spot. This is produced by some systematic offset of unknown origin, which only shows up for the thermal spot (full and magnetic-only sunspots do not present this issue). We have added seven seconds to the travel-time shift of the thermal sunspot for TD1 and frequency 4.25 mHz in order to account for this offset.  

Except for some combinations of low frequency and low phase-speed filters, most of the shifts are negative. The $p$ mode waves need less time to complete their travel in this model than in the quiet Sun. This fact contrasts with the sound speed perturbation of the sunspot model (Figure \ref{fig:sunspot_model}) which includes a reduction of the sound speed in a shallow layer near the surface. If the wave speed were the main contribution to the travel-time shifts, one would expect to obtain positive shifts, since the waves travel more slowly in the thermal sunspot than in the quiet Sun. However, our results show significant negative travel times, and point to another cause. The most plausible cause of this feature is the effect of the changes in density and sound speed on the cut-off frequency. The thermal spot modifies the height of the upper turning point of the waves with respect to the quiet Sun atmosphere. Since the reflection layer is shifted to lower heights, the path of the waves is shorter and, thus, the time needed to complete it is also shorter. This result will be discussed in more detail in Section \ref{sect:rays}.

\begin{figure}[!ht] 
 \centering
 \includegraphics[width=9cm]{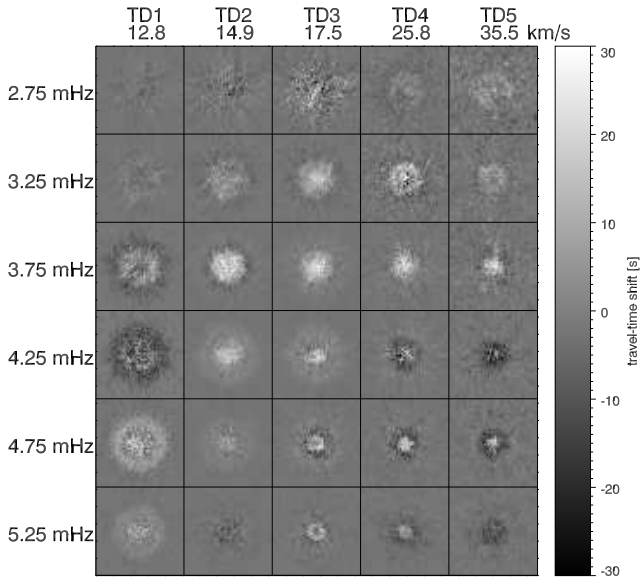}
  \caption{Mean travel-time shifts measured for the magnetic-only sunspot simulation using phase speed and and frequency bandpass filters, as in Figure \ref{fig:sunspot travel-time phase}.}
  \label{fig:magnetic travel-time phase}
\end{figure}

\subsection{Magnetic-only spot travel-time maps}
\label{sect:thermal travel-time}

Figure \ref{fig:magnetic travel-time phase} shows the mean travel-time shifts for the magnetic-only sunspot. This simulation includes the direct effect of the magnetic field on the wave propagation, but neglects the thermal variations of the atmosphere. Our measurements show mostly positive travel-time shifts. The strongest travel-time shifts are found for waves with frequencies around 3.75 mHz, which show shifts around 20 s. The magnitude of the signal decreases for higher frequencies and also for higher phase speeds, and some filter combinations show slightly negative travel-time shifts.

Obtaining longer travel times in a magnetized atmosphere seems
counterintuitive, since one would expect the fast magneto-acoustic
mode propagating faster in the regions where the Alfv\'en speed is
higher, leading to shorter travel times. However, travel-time
perturbations are sensitive to changes in the phase of the wave.
In magnetic fields, phase changes may be caused by a variety of processes
including mode conversion, transmission, and reflection of fast waves
\citep[\eg][]{Zhugzhda+Dzhalilov1982,Bogdan+etal2002,Cally2005,Cally2006,Khomenko+Collados2006}. Our results are consistent with \citet{Cally2009}, who  evaluated the travel-time perturbations produced by an uniform
magnetic field added to the quiet Sun Model S of \citet{Christensen-Dalsgaard+etal1996}. For a vertical magnetic field with 2500 G strength
(like the umbra of our model) he obtained mainly positive travel-time
shifts which are in qualitative agreement with our measurements.
In both analyses higher frequencies (around 5
mHz) show lower travel-time perturbations than waves with 4 mHz
frequency, and they even exhibit small negative travel-time shifts.
The effect of the magnetic field is also lower for deep propagating
waves (higher phase speed), as they reach the surface almost
vertically and, thus, interact only weakly with the vertical magnetic
field.

\subsection{Model comparison}

\begin{figure}[!ht] 
 \centering
 \includegraphics[width=9cm]{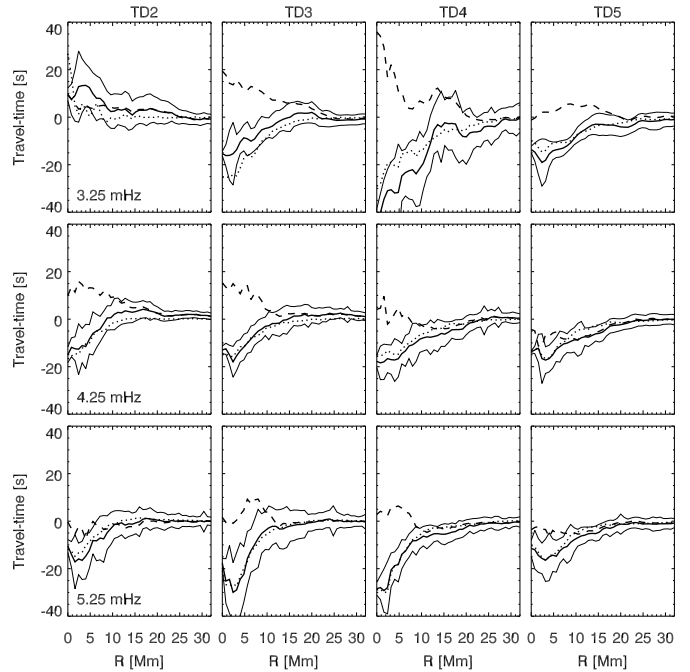}
  \caption{Azimuthal averages of the mean travel-time shifts measured for the full (solid line), thermal (dotted line), and magnetic-only (dashed line) sunspots. Thin solid lines enclose the region limited by the errors of the full sunspot. Columns from left to right illustrate phase speed filters from TD2 to TD5. Rows shows frequency filters centered at 3.25 mHz (top panels), 4.25 mHz (middle panels), and 5.25 mHz (bottom panels).}
  \label{fig:traveltime_average_TD}
\end{figure}

Figure \ref{fig:traveltime_average_TD} shows a comparison of the azimuthally averaged travel-time shifts obtained for the full, thermal, and magnetic-only sunspots. The errors were estimated as the standard deviation of the data values inside each annular region used for the average. Only the errors of the full sunspot measurements are illustrated in order to simplify the plots. The thermal sunspot shows similar errors, while the magnetic-only sunspot presents higher errors for certain combinations of phase speed and frequency filters. As noted earlier, waves with low phase speed and low frequency (\eg, phase speed filter TD2 and frequency filter 3.25 mHz) show positive travel-time shifts for the full and thermal sunspots. For low frequency waves with higher phase speed, there is a good qualitative agreement between the thermal and full sunspots, both show negative travel-time shifts with a signal between -20 and -40 s. Travel-time perturbations measured with the 4.25 mHz and higher frequency filters show a remarkable quantitative agreement between thermal and full sunspots. This agreement is strikingly good for high phase speed filters, where the thermal sunspot seems to provide an accurate characterization of the full sunspot travel-time measurements. It must be noted that the signals obtained from the magnetic-only sunspot simulation are completely different from those measured for the other numerical experiments. The magnetic-only sunspot produces mainly positive travel-time shifts, showing in most cases an opposite sign than that of the models which include the thermal perturbation.

\section{Ray approximation estimates}
\label{sect:rays}

The similarities between the travel-time shifts measured for the full sunspot and the thermal sunspot strongly suggest that the thermal contribution is the main effect observed in travel-times, at least for a large subset of phase speed and frequency filters. This conclusion is reinforced by the fact that the magnetic-only sunspot produces travel-time shifts with opposite signs. However, the negative travel-time shifts obtained for the thermal and full sunspots indicate that the signal is not caused by changes in the wave speed (the sound speed is reduced) but it is a consequence of the Wilson depression instead. To better illustrate these effects, we have compared the holography results retrieved from the numerical simulations with some simple ray approximation calculations following \citet{Lindsey+etal2010}.

The depth of the upper turning point of the wave depends on the frequency according to the dispersion relation and the depth dependence of the acoustic cut-off frequency. In this form of the ray approximation, dispersion is included and the phase travel time for waves with vertical incidence is estimated as

\begin{equation}
t =\frac{1}{\omega}\int_{z_0}^{z(\tau=0.01)}{kdz},
\label{eq:travel_time}
\end{equation}

\noindent where $\omega$ is the frequency, $z(\tau=0.01)$ is the atmospheric height where the oscillations are measured (corresponding to the height with optical depth $\tau=0.01$ in our analysis), and $z_0$ is some reference depth below the photosphere sampled by the seismic observation. The wavenumber is given by the dispersion relation for vertically propagating waves:

\begin{equation}
k^2 =\frac{\omega^2-\omega_{c}^2}{c^2},
\label{eq:dispersion_relation}
\end{equation}

\noindent where the expression for the acoustic cut-off frequency \citep{Lindsey+etal2010} is

\begin{equation}
\omega_{c}=\frac{c}{2H}\Big (1+2\frac{dH}{dz}\Big)^{\frac{1}{2}}
\label{eq:cut_off}
\end{equation}

\begin{figure}[!ht] 
 \centering
 \includegraphics[width=9cm]{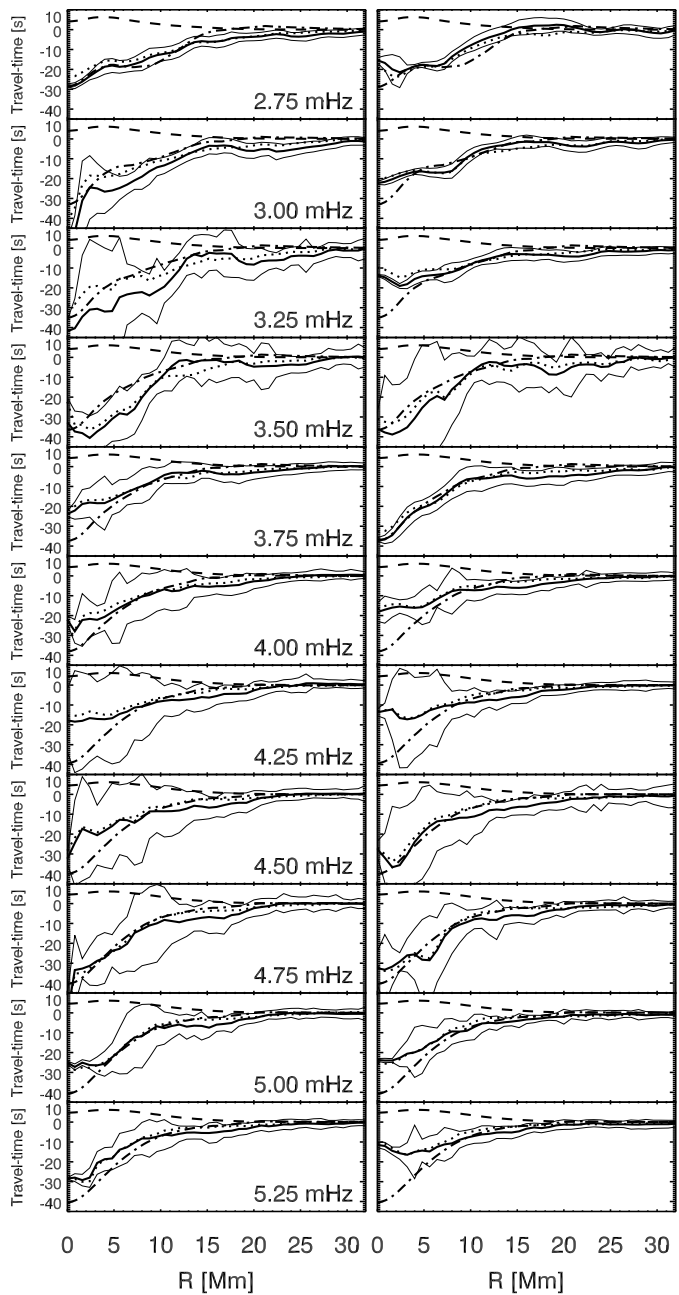}
  \caption{Azimuthal averages of the mean travel-time shifts measured for the full (solid line) and thermal (dotted line) sunspot, and estimated by the ray approximation for the (dash-dotted line) for the phase speed filters TD4 (left panels) and TD5 (right panels). Thin solid lines enclose the region limited by the errors of the full sunspot. Dash lines illustrate the contribution of the sound-speed perturbation. Rows from top to bottom illustrate different frequency filters from 2.75 mHz to 5.25 mHz.}
  \label{fig:traveltime_average_lindsey2010}
\end{figure}

\noindent In these expressions $c$ is the sound speed and $H$ is the density scale height. The travel-time shift is given by the difference of the travel times $t$ between the quiet-Sun and sunspot models.  Since we are isolating the effects of the Wilson depression, we are neglecting the direct effects of the magnetic field which might be important at lower depths. For vertically propagating waves the contribution of the buoyancy to Equation \ref{eq:dispersion_relation} also vanishes. Equation \ref{eq:travel_time} is integrated up to the upper turning point, which is typically deeper than $z(\tau=0.01)$. The depth $z_0$ was chosen at -20 Mm, deep enough to include the perturbation inside the integration range. Tests were computed with several values for $z_0$ which showed that this selection does not matter as long as $z_0$ is deeper than $z=-2$ Mm. Figure \ref{fig:traveltime_average_lindsey2010} shows a comparison between the thermal and full sunspot travel-time shifts obtained from the holography measurement and the ray approximation for the phase speed filters TD4 and TD5. We have chosen to only include the two filters with higher phase speeds because these waves propagate to deeper layers and their incidence at photospheric heights is more vertical than for lower phase speed waves. The comparison of the other phase speed filters would require a full ray calculation, but this is out of the scope of this paper.   

Figure \ref{fig:traveltime_average_lindsey2010} shows the variation of the travel-time shift with the radial position for all the frequencies analyzed in this study. The travel-time shift at each radial position was obtained by integrating Equation \ref{eq:travel_time} at the corresponding horizontal position in the sunspot model. There are some evident differences between the holography measurements and the ray computations. These discrepancies are particularly obvious near the center of the spot for some frequencies, \ie, for both phase speed filters at 4.00 and 4.25 mHz and for the TD5 filters above 5.00 mHz.  In these cases, the differences between both measurements are above the estimated errors. However, the ray calculation quantitatively reproduces the holography signals for the rest of the measurements. For all frequencies, a noticeable agreement is found for radial positions higher than 5 Mm, while certain frequencies such as 2.75, 3.75, or 4.75 show a striking match between both results. It must be noticed that this simple implementation of the ray approximation does not account for all the effects expected from detailed wave mechanics, but still this comparison points out that the changes in the reflection depth produced by the Wilson depression are a fundamental feature for the interpretation of travel-time shifts in sunspots.

Helioseismic holography can capture not only the effect of the Wilson depression on the travel times, but also the travel-time shift produced by changes in the sound speed. Previous studies have performed holography analyses of acoustic wave propagation through models with a perturbation in the sound speed, proving that the inversion of the measured travel-time shifts is able to recover the sound speed target \citep[\eg,][]{Birch+etal2009}. Although the travel-time shifts obtained in our sunspot are likely caused by the Wilson depression, it is interesting to quantify the contribution of the strong sound-speed variations (see Figure \ref{fig:sunspot_model}) to the measured travel-time anomalies. We have addressed this issue by isolating the sound-speed perturbation effects on a vertical ray calculation. The results discussed in the previous paragraphs account for the variations in the sound speed and Wilson depression. We have evaluated the travel times in the quiet Sun atmosphere (with quiet Sun sound speed) of vertical rays between the height $z_0$ and the geometrical position of the upper turning point of the sunspot model (instead of the quiet Sun). That is, the path of this measurement is equal to that from the sunspot estimation. The difference between the sunspot and this new calculation is illustrated by the dashed line in Figure \ref{fig:traveltime_average_lindsey2010}. Since the sound speed of the sunspot is lower than the quiet Sun values, its contribution to the total travel-time shift is positive. However, the ray is reflected before reaching the region of the model with strongest sound-speed perturbation. The sound-speed perturbation produces an increase in the travel time, opposite to the effect of the Wilson depression, with a maximum contribution around 6 s. This contribution is almost independent of frequency.

\section{Discussion and conclusions}
\label{sect:conclusions}

The direct effects of the magnetic field are expected to play a key role in the wave propagation through magnetized atmospheres and, thus, including the full MHD must be the next big step in the interpretation of the helioseismic signals in order to achieve a more reliable characterization of the atmosphere below sunspots. The travel-time shifts obtained from our numerical simulations of wave propagation in a MHS sunspot model capture most of the properties previously measured in sunspot observations. Waves with low phase speed show positive travel-times (they seem to propagate slower), while waves with high phase speed present negative travel-times (they seem to propagate faster). Since the depth of the penetration of the modes increases with the phase speed, the interpretation of those measurements as a consequence of variations in the sub-surface wave speed produces a two-layer model \citep{Kosovichev+etal2000, Couvidat+etal2005}, which extends down to approximately 10 Mm below the surface. However, the sound speed in our sunspot model only has a near surface reduction, so the assumption that the travel-time shifts are a result of perturbations in the wave speed cannot explain the measurements.

We have further explored this phenomenon by analyzing a numerical simulation where only the thermal contribution of the sunspot is included, while the magnetic field is set to zero. The travel-time shifts show negative values, even though the model only has a reduction in the sound speed. This indicates that the changes that the thermal effects produce on the acoustic cut-off frequency have a stronger influence on the travel time than the wave speed. These results and conclusion agree with \citet{Moradi+etal2009}. In the following sections we discuss the results for different phase speed filters:

\subsection{High phase speed filters (TD4 and TD5)}

The two highest phase speed filters analyzed in this work show negative travel-time shifts for all the frequencies studied, from 2.75 mHz to 5.25 mHz. These waves propagate to deeper layers (their lower turning point is between $z=-6$ Mm and $z=-10$ Mm) and when they return towards the surface their angle of incidence is almost vertical. The interaction of these waves with the near-vertical umbral magnetic field is weak \citep{Cally2009, Lindsey+etal2010} and the travel-time shifts measured for the full sunspot are similar to those obtained for the thermal sunspot. This behavior is seen for all frequencies (see Figure \ref{fig:traveltime_average_lindsey2010}). The travel-time shifts of these two phase speed filters were evaluated in light of the estimations computed in the ray approximation for vertically propagating waves. The analyses indicate that the travel-times are mainly affected by changes in the wave path rather than variations of the wave speed.  The perturbation of the thermodynamic variables (density and sound speed) associated to the presence of the sunspot magnetic field produces changes in the cut-off frequency. In the sunspot atmosphere the location of the upper turning point of the waves is deeper, and the path that the waves follow is shorter. The travel times of the waves is also reduced, generating the negative travel-time shifts that traditionally have incorrectly been interpreted as an increase in the wave speed. Thermal effects are the main cause of the measured travel-time shifts. It must be noted that the comparison shows some significant discrepancies for certain frequencies near the center of the spot. A more detailed (finite wavelength) calculation would be necessary to address this mismatch. 

A thorough comparison between the radial variation of the full and thermal sunspots travel-time shifts reveals that, while the agreement for radial distances within 10 Mm is excellent, at radial distances between 10 and 20 Mm they slightly depart from each other. Generally, the magnitude of the full sunspot travel-time shift is greater than that of the thermal sunspot. Although the differences between both measurements are inside the error limits, they are consistent for several frequencies in the range between 4.25 and 5.25 mHz. In addition, it is interesting to note that in that region the ray calculation matches the thermal travel-time shift but departs from the full sunspot. A plausible cause of this is the direct effect of the magnetic field in the travel-time. Mode conversion from fast acoustic to fast magnetic waves is more efficient when the angle between the wave vector and the magnetic field is higher \citep{Cally2005} and, thus, the magnetic effects in the inclined penumbral field are expected to be more important. When the converted atmospheric fast magnetic wave is reflected back towards the interior due to the gradients in the Alfv\'en speed, it makes a contribution to the travel-time shift that is not experienced by the acoustic wave in the thermal sunspot model.

\subsection{Medium phase speed filters (TD2 and TD3)}

The travel-time shift measured with these phase speed filters changes from positive values at low frequencies to negative values at higher frequencies. The frequency at which this change occurs depends on the phase speed; it happens at 3.75 mHz for TD2 and at 3.25 mHz for TD3. Above these frequencies, the behavior of the measured travel-time shift for these phase speed filters is similar to higher phase speeds: the negative travel-time shifts of the full sunspot show a striking agreement with the thermal sunspot. For these combinations of phase speed and frequency filters the direct magnetic field effects also are small. 

On the other hand, the magnitude of the full sunspot travel-time shift of the TD2 filter between 3.50 and 4.00 mHz at the center of the spot shows significantly lower values than the thermal sunspot. The travel-time shift of the thermal sunspot for these frequencies is between -30 and -40 s, while the full sunspot travel-time shift goes from approximately 0 at 3.50 mHz to -15 s at 4.00 mHz. In addition, the thermal sunspot measurements show a remarkable agreement with the ray approximation calculation (this comparison is not plotted in the figures). This behavior can also be interpreted in terms of the dependence of the interaction between fast waves and magnetic field with the angle of incidence. Lower phase speed waves propagate in a shallower cavity and when the ray reaches the surface it forms a large angle with the vertical umbral magnetic field. Mode conversion is efficient under these conditions and direct magnetic effects are evident. However, it must be noticed that the magnetic-only sunspot cannot reproduce the full sunspot travel-time shift (compare the travel-time shift at TD2 and 3.75 mHz from Figures \ref{fig:sunspot travel-time phase} and \ref{fig:magnetic travel-time phase}). Mode conversion occurs at the height where the sound and the Alfv\'en speeds are similar. Since the magnetic-only sunspot has a quiet Sun thermal structure, the height of the conversion layer differs in both simulations and the wave ray hits this layer at a different angle. Thus, the efficiency of mode conversion in both cases is different, leading to a disagreement in the travel-time shift. The values of the full sunspot travel-time shifts are between the magnetic-only and thermal sunspots measurements.

Below 3 mHz the full sunspot and the thermal sunspot show remarkable agreement. This is in sharp contrast with  the results found for frequencies between 3 and 4 mHz, where the full and thermal sunspots travel-time shifts disagree. For low frequencies magnetic effects are less relevant, which is consistent with the lower efficiency of mode conversion for those waves \citep{Cally2005}. In addition, these low frequencies show a positive travel-time shift. The fact that the thermal case shows a sign change with frequency indicates that the sensitivity of the sign of the travel-time shift is not a magnetic effect. Instead, this property might depend on the choice of data analysis filters, as suggested by \citet{Braun+Birch2008} and \citet{Birch+etal2009}. This effect, in principle, can be understood and included in modeling efforts.

Figure \ref{fig:sunspot travel-time phase} shows an annular pattern of reversed travel-time shift for phase speed filters TD2 and TD3 at certain frequencies. This feature shows similarities with the ring-like regions  identified observationally by \citet{Couvidat+Rajaguru2007} for the same phase-speed filters. Their measurements showed a positive travel-time shift surrounded by a negative travel-time shift, although it must be noticed that their analysis was not filtered in frequency and it is not directly comparable to our results. The fact that this annular pattern vanishes in the thermal sunspot (Figure \ref{fig:thermal travel-time phase}) points to the effect of the magnetic field in the generation of this feature, as suggested by \citet{Couvidat+Rajaguru2007}, since surface effects may be enhanced in regions with inclined magnetic fields. However, our results cannot discard other causes that may lead to differences between spot center and edges (\eg, nonlinearities combined with filter effects).

\subsection{Low phase speed filter (TD1)}

The phase-speed filter TD1 shows a complex behavior and its interpretation is less straightforward than the higher phase speed filters. At high frequencies the full sunspot presents negative travel-time shifts, in quantitative agreement with the thermal sunspot. This behavior is similar to the other phase speed filters. 

For frequencies between 3.75 and 4.75 mHz the travel-time shifts
change from positive to negative. Although the thermal sunspot also
shows a change in the sign of the travel-time, its values disagree
with those from the full sunspot. Several causes can explain a higher
sensitivity of the TD1 filter to the magnetic field. First, low phase speed waves travel closer to the surface, where magnetic field is
significant, and spend more time near the conversion layer; second,
their angle of incidence is more horizontal, forming a larger angle
with the strong nearly vertical magnetic field around the center of
the sunspot. The positive travel-time shifts measured for TD1 in low
frequencies are opposite to what one would expect for shorter travel
paths due to the Wilson depression. Since the thermal model also shows
these positive travel-time shifts, it is clear that they are not only
due to magnetic effects. In addition to the higher magnetic
sensitivity, it is also known that results using filters at low phase
speeds (such as TD1) are highly sensitive to changes of the filter width
and central phase-speed \citep{Braun+Birch2008, Moradi+etal2009}.
This sensitivity can even include sign reversals of the travel-time shift.
\citet{Braun+Birch2008} determined that the cause of this behavior
is the background power between the $p_1$ and $f$ ridges. Thus, the
interpretation of the TD1 measurements is not straightforward. As seen in Figure \ref{fig:sunspot travel-time phase}, the travel-time shift maps of the phase speed filter TD1 show some remarkable rings with reversed sign. These rings may be related to magnetic field effects or may represent an artifact of the analysis methods.

\subsection{Final remarks}

The main conclusion extracted from our numerical simulations is that the thermal component of the sunspot provides an accurate characterization of the travel-time shifts produced by the full sunspot for some combinations of phase speed and frequency filters (e.g. filters TD4, TD5, and the high frequency - low phase-speed regime). This opens a new opportunity to inversion methods of travel-time perturbations. Including the direct effects of the magnetic field in those inversions has proven to be a formidable task that despite the efforts of the helioseismic community still remains incomplete. Our results suggest some progress may be achieved by choosing the appropriate combination of phase speed and frequency filters and including the effect of the Wilson depression, in addition to the sound speed, in the inversion. This approach would allow bypassing the inclusion of direct effects of the magnetic field to which the selected waves are rather insensitive. However, some caution must be taken. First, inclined magnetic fields at the penumbra can modify the upper turning point since they provide portals for low frequency waves \citep{Jefferies+etal2006}. Magneto-acoustic waves with frequencies ($\omega$) below the acoustic cut-off ($\omega_c$) can propagate upwards if $\omega >\omega_c\cos\theta$, where $\theta$ is the inclination of the magnetic field. The change of the path produced by this magnetic modification of the acoustic cut-off may also leave a trace in the travel-time perturbation. In addition, the reflected atmospheric fast and downgoing Alfv\'en waves in active regions can return to the interior and contribute to the helioseismic signal \citep{Cally+Moradi2013}. Due to the limited height of the top boundary in our simulations, these waves might not have enough room to complete their reflection, and part of their signal may be lost in our analysis.

Finally, it should be emphasized that our conclusions are based of the analysis of a specific sunspot model that only resembles qualitatively the measurements obtained from actual observations. Our interpretations should be validated by more detailed analyses, including more sophisticated ray tracing or improved numerical models. In order to fully characterize the travel times measured in typical sunspots, those models should address all the influences which have been proposed to have an effect on the measurements: magnetic field strength and inclination \citep{Schunker+etal2005, Cally2009}, Wilson depression \citep{Lindsey+etal2010, Schunker+etal2013}, sound speed perturbations \citep{Kosovichev1996, Couvidat+Rajaguru2007}, flows \citep{Duvall+etal1996}, wave amplitudes \citep{Rajaguru+etal2006}, or multiple scattering by bundled flux tubes \citep{Felipe+etal2013, Felipe+etal2014, Hanson+Cally2014}. Inferring the structure of sunspots accounting for all these effects remains a challenge for local helioseismology.

\acknowledgements This research has been funded by the Spanish MINECO through grant AYA2014-55078-P. At NWRA, support for this work is provided by the NASA Living With a Star program through grant NNX14AD42G, by the NASA Heliophysics Supporting Research program through contract NNH12CF23C, and by the Solar Terrestrial program of the National Science Foundation through grant AGS-1127327. ACB acknowledges the EU FP7 Collaborative Project ``Exploitation of Space Data for Innovative Helio- and Asteroseismology'' (SPACEINN). This work used the NASA's Pleiades supercomputer at Ames Research Center, MareNostrum supercomputer at Barcelona Supercomputing Center, and Teide High-Performance Computing facilities at Instituto Tecnol\'ogico y de Energ\'ias Renovables (ITER, SA).

\end{document}